# Multigrid Monte Carlo in the Sine Gordon model[*]

Martin Grabenstein and Bernhard Mikeska

II. Institut für Theoretische Physik, Universität Hamburg,
Luruper Chaussee 149, 22761 Hamburg, Germany

We pose two questions about the dynamical critical behavior of multigrid Monte Carlo: Will a multigrid Monte Carlo simulation of the two dimensional Sine Gordon model exhibit critical slowing down, as expected by a theoretical analysis of Metropolis acceptance rates? Can we reduce critical slowing down caused by decreasing acceptance rates on large blocks by performing more updates on coarser lattices? To this end we simulate the model with a W-cycle ($\gamma = 2$) and a higher cycle with $\gamma = 4$ using piecewise constant interpolation. The answer to the first question is positive, the answer to the second one is negative.

## 1. MOTIVATION

Critical slowing down (CSD) causes severe limitations in computer simulations with local algorithms in statistical mechanics and lattice gauge theory close to a critical point. The multigrid Monte Carlo approach [1] is an attempt to overcome this problem by using a nonlocal updating scheme.

A step towards a better insight *why* multigrid Monte Carlo is successful in beating CSD for some models while it does not work as well for others, was made in [2]: An analytic calculation and analysis of acceptance rates for nonlocal Metropolis updating $\phi_x \to \phi_x + s\psi_x$ was performed. $s$ is a random number, e.g. equally distributed in the interval $[-\varepsilon, \varepsilon]$, and $\psi_x$ denotes a suitably normalized coarse-to-fine interpolation kernel. The simplest example for $\psi_x$ is a piecewise constant kernel: $\psi_x = const$ inside a given block, and $\psi_x = 0$ outside. It was argued that for a critical model with Hamiltonian $\mathcal{H}(\phi)$, CSD will occur if the expansion of $\langle \mathcal{H}(\phi + \psi) \rangle$ in terms of $\psi$ contains a relevant (mass) term $\sim \sum_x \psi_x^2$. If such a term is present, Metropolis step sizes $\varepsilon(L_B)$ on block lattices with increasing block size $L_B$ have to be scaled down like $\varepsilon(L_B) \sim L_B^{-1}$ in order to obtain block size independent acceptance rates. This strong decrease of step sizes on large blocks was found in several models for smooth and piecewise constant interpolation. One of these models is the Sine Gordon model in two dimensions.

As the first question to be answered here, we want to check the prediction [2] that a W-cycle (cycle control parameter $\gamma = 2$) with piecewise constant interpolation will not eliminate CSD in the rough (massless) phase of the Sine Gordon model [3]. (This algorithm eliminates CSD in the Gaussian model.)

## 2. THE MODEL

The 2-d Sine Gordon model is defined on an $L \times L$ lattice $\Lambda_0$ via the partition function

$$Z = \int \prod_{x \in \Lambda_0} d\phi_x \exp(-\mathcal{H}(\phi)) , \qquad (1)$$

with the Hamiltonian

$$\mathcal{H}(\phi) = \frac{1}{2\beta} \sum_{\langle x,y \rangle} (\phi_x - \phi_y)^2 - \zeta \sum_x \cos 2\pi \phi_x . \qquad (2)$$

From the point of view of statistical mechanics, this system can be considered as a 2-d surface in a periodic potential. The model exhibits a Kosterlitz-Thouless phase transition at $\beta_c(\zeta)$. In the limit of vanishing fugacity $\zeta$, $\beta_c$ takes the value $2/\pi = 0.6366\ldots$. For $\beta > \beta_c$ the model is in the rough (massless) phase. There, the cosine-term of the Hamiltonian is irrelevant in the renormalization group sense. The system is critical and has the same long distance behavior as the

---

[*]To appear in the Proceedings of LATTICE 93, Dallas, USA, October 1993. Talk presented by M. Grabenstein.

Table 1
Numerical results for the 2-d Sine Gordon model on $L \times L$ lattices in the rough phase, $\beta = 1.0$, $\zeta = 0.5$.

| | $\gamma = 2$ | | | $\gamma = 4$ | | |
|---|---|---|---|---|---|---|
| $L$ | statistics | $\sigma^2$ | $\tau_{int,\sigma^2}$ | statistics | $\sigma^2$ | $\tau_{int,\sigma^2}$ |
| 4   | 25 000  | 0.268(1)   | 0.96(3)    | 25 000  | 0.268(1)   | 0.91(3)   |
| 8   | 50 000  | 0.3809(9)  | 1.35(3)    | 25 000  | 0.380(1)   | 1.14(4)   |
| 16  | 100 000 | 0.4896(7)  | 2.70(8)    | 25 000  | 0.488(1)   | 1.67(6)   |
| 32  | 300 000 | 0.5996(7)  | 8.54(19)   | 100 000 | 0.5997(9)  | 4.15(11)  |
| 64  | 500 000 | 0.7105(10) | 30.5(1.0)  | 300 000 | 0.7113(9)  | 14.2(4)   |
| 128 | 500 000 | 0.8218(19) | 113.7(6.9) | 300 000 | 0.8213(18) | 58.2(3.3) |

massless Gaussian model. The fluctuations of the surface are given by the surface thickness

$$\sigma^2 = \langle (\phi_x - \overline{\phi})^2 \rangle \;, \qquad (3)$$

where $\overline{\phi}$ denotes the average of the field over the lattice. In the rough phase, the surface thickness $\sigma^2$ scales with $\log L$ [4].

### 3. SIMULATION WITH A W-CYCLE

Our simulations are organized as follows [3]: We use a recursive multigrid algorithm, piecewise constant interpolation and a staggered coarsening with a factor of two as described in [5]. As pre-smoothing and post-smoothing operations, we choose a sweep of single hit Metropolis updates. The maximum Metropolis step size $\varepsilon(L_B)$ is scaled down like $L_B^{-1}$. Then, acceptance rates of approximately 50% are observed on all block lattices, in accordance with the theoretical analysis of [2]. As multigrid cycle we use a W-cycle with cycle control parameter $\gamma = 2$. The rule for multigrid cycles is that from an intermediate block lattice one proceeds $\gamma$ times to the next coarser lattice before going back to the next finer lattice. In this way, $\gamma$ times more updates on each coarser lattice are performed.

To study the dynamical critical behavior of the algorithm, we perform the simulation at $\beta = 1.0$, $\zeta = 0.5$, which is deep in the rough phase. There the correlation length is infinite and the physical length scale is set by the linear size of the lattice $L$. Thus, we expect the autocorrelation time $\tau$ to diverge with the dynamical critical exponent $z$ like $\tau \sim L^z$.

After equilibration, measurements were taken at each visit of the finest lattice. We computed the integrated autocorrelation times and the corresponding errors by a window method [6] with a self-consistent truncation window of $4\tau_{int}$ for the surface thickness $\sigma^2$. We checked that the autocorrelation functions showed an exponential decay.

The numerical results are given in Table 1. The autocorrelation times $\tau_{int}$ are measured in the number of visits on the finest lattice. Note that our runs are longer than 10 000 $\tau_{int}$ (longer than 4 000 $\tau_{int}$ on the $128^2$ lattice). Fig.1 shows the dependence of $\tau_{int,\sigma^2}$ on $L$ for the W-cycle ($\gamma = 2$). By fitting our data in the range $32 \leq L \leq 128$ with the ansatz $\tau_{int,\sigma^2} = cL^z$, we estimate $z = 1.9(1)$ for the W-cycle ($\gamma = 2$). Thus, as already predicted in [2], CSD in the rough phase of the Sine Gordon model is not reduced by a W-cycle with piecewise constant interpolation. According to the acceptance analysis, we would expect the same result with smooth interpolation.

### 4. HIGHER CYCLE ALGORITHM

The second question we want to address is whether one can circumvent slowing down caused by too small steps on large blocks by accumulating many of these steps randomly.

A constant accumulated step size on all length scales could be achieved in the following way: For step sizes scaling down like $\varepsilon(L_B) \sim L_B^{-1}$ and a coarsening by a factor of two, the Metropolis step size on a next coarser grid is too small by a factor of two. If we assume that subsequent update steps within a multigrid cycle are independent and accumulate in a random-walk like way, we can expect to compensate for this decrease of step sizes by increasing the number of updates on the next coarser grid by a factor of four. This can be achieved by a higher cycle with cycle control parameter $\gamma = 4$. An example for such a multigrid cycle is given in Fig.2. For this cycle, four times more updates on each coarser lattice are performed. For a recursive multigrid algorithm, the computational effort is $\sim L^d$ for $\gamma < 2^d$ and $\sim L^d \log L$ for $\gamma = 2^d$ in $d$ dimensions [5]. Therefore a higher cycle with $\gamma = 4$ is practical for $d > 2$ and borderline practical for $d = 2$. In summary, if the random-walk picture would be correct, the acceptance problem could be solved by doing more work on coarser lattices.

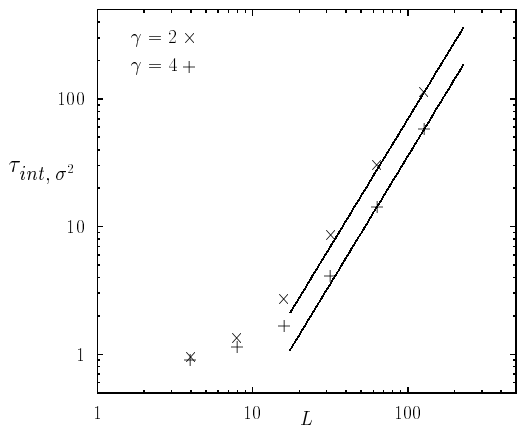

Figure 1. Dependence of the integrated autocorrelation time for the surface thickness $\sigma^2$ on the lattice size $L$ in the rough phase of the 2-d Sine Gordon model, $\beta = 1.0$, $\zeta = 0.5$. Errors are smaller than the symbols used. The lines correspond to $z = 2$.

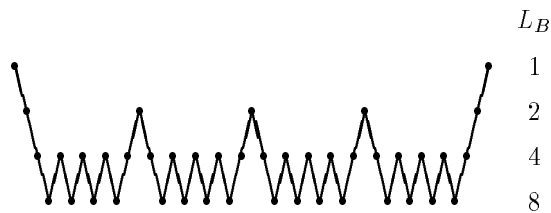

Figure 2. Higher cycle with $\gamma = 4$

To investigate this possibility we repeated our runs, now changing from the W-cycle ($\gamma = 2$) to the higher cycle with $\gamma = 4$. Our results for the higher cycle are also contained in Table 1 and plotted in Fig.1. Analogous fits as above yield $z = 1.9(1)$ for the higher cycle ($\gamma = 4$), which is the same value as the exponent for the W-cycle ($\gamma = 2$). The results clearly show that compensating for decreasing acceptance rates on large blocks by choosing a higher cycle with $\gamma = 4$ does not improve the dynamical critical behavior of the algorithm. We conclude that a random-walk like argumentation as stated above is not correct in the case of the Sine Gordon model.

We would like to thank G. Mack, S. Meyer and K. Pinn for helpful discussions. This work was supported in parts by the Deutsche Forschungsgemeinschaft, the German Israeli Foundation and the German Scholarship Foundation. Our simulations were performed on hp 9000/730 RISC workstations at DESY.